# Electric Vehicle Routing Problem with Time Windows and Station-based or Route-based Charging Options

Tran Trung Duc[1,2,6], Vu Duc Minh[3,4] ✉, Nguyen Ngoc Doanh[1,2], Pham Gia Nguyen[3,7], Laurent El Ghaoui[1,2], Ha Minh Hoang[3,4,5]

[1] VinUniversity, Vietnam [2] Center for Environmental Intelligence, VinUniversity [3] Smart Logistics and Supply Chain Management Laboratory [4] Center for Applied Data Science and Artificial Intelligence [5] National Economics University, Vietnam [6] CMC University, Vietnam [7] Hanoi University of Science, Vietnam
✉ Corresponding Author: minhvd@neu.edu.vn

**Abstract.** The Electric Vehicle Routing Problem with Time Windows and Station-based or Route-based Charging Options addresses fleet optimization incorporating both conventional charging stations and continuous wireless charging infrastructure. This paper extends Schneider et al.'s foundational EVRP-TW model with arc-based dynamic wireless charging representation, partial coverage modeling through parameter $\omega_{ij}$, and hierarchical multi-objective optimization prioritizing fleet minimization. Computational experiments on Schneider benchmark instances demonstrate substantial operational benefits, with distance and time improvements ranging from 0.7% to 35.9% in secondary objective components. Analysis reveals that 20% wireless coverage achieves immediate benefits, while 60% coverage delivers optimal performance across all test instances for infrastructure investment decisions.

**Keywords:** Electric vehicle routing problem, dynamic wireless charging, time windows, mixed integer programming, fleet optimization.

## 1    Introduction

Transportation sector accounts for approximately 25% of global $CO_2$ emissions, with road transportation representing 75% of these emissions [1]. The rapid growth of e-commerce and last-mile delivery services has intensified this environmental impact, making sustainable transportation solutions increasingly critical [2]. Electric vehicles (EVs) offer zero local emissions and superior efficiency in urban environments [3].

The classical Vehicle Routing Problem (VRP), first introduced by Dantzig and Ramser [4], has been extensively studied for optimizing delivery routes while minimizing transportation costs. However, the unique characteristics of electric vehicles, particularly their limited driving range and charging requirements, necessitated fundamental extensions to traditional VRP formulations. This family of problems, comprehensively reviewed by Erdelić and Carić [5], extends classical VRP by explicitly incorporating energy consumption constraints and charging infrastructure considerations.

Schneider et al. [6] established the foundational mathematical framework for the Electric Vehicle Routing Problem with Time Windows and Recharging Stations (EVRP-TW), extending classical VRP to include battery capacity constraints and charging station visits. Their work introduced the fundamental mathematical structure for battery state tracking, energy consumption modeling, and discrete charging station integration within routing decisions, assuming vehicles must visit designated charging stations and remain stationary during battery replenishment.

Recent technological advances in Dynamic Wireless Charging (DWC) systems enable continuous energy transfer to moving vehicles through electromagnetic induction from infrastructure embedded in roadways, potentially eliminating charging downtime and extending operational range [7]. Feasibility studies demonstrate that DWC can reduce battery capacity requirements by 27–44% while decreasing vehicle weight by 12–16% and energy consumption by 5.4–7% [8].

Despite these technological advances, existing EVRP mathematical formulations have not adequately incorporated wireless charging capabilities. Current models suffer from binary charging availability assumptions that ignore partial wireless coverage reality, inadequate integration with established frameworks, and lack of comprehensive cost optimization considering both fleet utilization and operational expenses.

To bridge this research gap, this paper introduces a comprehensive mathematical framework for the Electric Vehicle Routing Problem with Time Windows featuring mixed charging options, designated as EVRPTW-DWC (Dynamic Wireless Charging). Our framework builds upon Schneider et al.'s foundational work [6] by integrating strategic on-route dynamic wireless charging, delivering substantial cost and time advantages over traditional station-based charging methods.

The remainder of this paper is organized as follows: Section 2 provides a literature review tracing the evolution from Schneider et al.'s foundational EVRP-TW model to current wireless charging developments. Section 3 presents the mathematical model formulation with detailed notation and constraints. Section 4 discusses computational experiments and results analysis. Section 5 concludes with key findings and future research directions.

## 2   Literature Review

The growing adoption of electric vehicles (EVs) in logistics has spurred significant research into adapting classical vehicle routing models to accommodate energy-related constraints. Among these, the Electric Vehicle Routing Problem (EVRP) has emerged as a central focus, evolving from static charging strategies to more advanced dynamic wireless charging (DWC) approaches. This section reviews the progression of EVRP research, starting from foundational formulations, moving through enhancements that improve model realism, and culminating in the integration of dynamic charging mechanisms. The discussion highlights key contributions, model innovations, and unresolved challenges that inform the development of our proposed framework.

## 2.1 Foundational Electric Vehicle Routing Problem

Schneider et al. [6] introduced the first comprehensive mathematical formulation for the Electric Vehicle Routing Problem with Time Windows and Recharging Stations, establishing the foundational framework that has guided subsequent EVRP research. Their model extends the classical VRPTW by formalizing the core operational aspects of EVs, including mechanisms to track the battery's state of charge, manage visits to recharging stations, and adhere to customer time windows. This formulation established the mathematical foundation for modeling battery capacity constraints, linear energy consumption rates, and discrete charging station visits within routing optimization.

Following Schneider et al.'s foundational work, researchers have developed numerous extensions while maintaining the core mathematical structure. Keskin and Çatay [9] extended the model to allow the partial charging constraint, enabling vehicles to charge less than full capacity. Montoya et al. [10] developed nonlinear charging functions to more accurately represent battery charging behavior, while Goeke and Schneider [11] incorporated speed-dependent consumption models. These extensions improved model realism but retained the fundamental structure of station-based charging.

## 2.2 Evolution Toward Dynamic Charging Concepts

The first attempts to incorporate dynamic charging into EVRP formulations emerged in the late 2010s, building upon advances in wireless power transfer technology. Li et al. [12] made one of the first attempts by introducing a hybrid formulation that combined static stations with the concept of continuous energy transfer on certain routes. This added operational flexibility compared to the full-recharge assumption. This early formulation assumed constant wireless power transfer and efficiency across all road segments, providing a first step toward continuous charging modeling but oversimplifying real-world charging dynamics. refined this idea by developing models with more sophisticated, location-dependent charging availability, where only specific road segments were designated as wireless-enabled. This approach introduced the concept of wireless-enabled road segments but employed binary representation that fails to capture partial coverage scenarios characteristic of real DWC implementations.

Recent developments have shown promising results. Cao et al. [13] developed routing and charging optimization with dynamic electricity pricing, achieving 50.01% reduction in algorithm runtime with only 5.56% increase in charging costs. Akbari et al. [14] introduced the Electric Vehicle Traveling Salesman Problem with DWC, finding that 30–40% wireless coverage can extend EV driving range from 200 to over 2000 kilometers without mid-journey charging stops. Sayarshad [15] and Bhatti et al. [16] demonstrated cost reductions of up to 30% and 35%, respectively, through optimized EV routing and charging strategies, while maintaining stable energy levels within 10% and ± 5% of target values, respectively.

### 2.3  Research Gaps and Model Requirements

Despite recent advances, current dynamic charging EVRP models exhibit several critical limitations that prevent their practical implementation. Most existing models treat road segments as uniform charging zones, failing to capture the spatial variation in charging efficiency and partial coverage that characterizes real DWC deployments. Current models use binary representation of charging availability that oversimplifies the reality where DWC infrastructure covers only portions of road segments with varying efficiency. Many dynamic charging models develop entirely new formulations rather than extending proven frameworks like Schneider et al.'s model, losing the computational efficiency and theoretical robustness of established approaches.

This paper addresses the identified research gaps by extending Schneider et al.'s foundational EVRP-TW framework to incorporate comprehensive dynamic wireless charging capabilities. The proposed EVRPTW-DWC model maintains the computational efficiency and theoretical rigor of Schneider et al.'s framework while addressing current limitations through arc-based charging representation that captures continuous energy transfer along road segments, partial coverage modeling with parameter $\omega_{ij}$ representing realistic DWC deployment scenarios, hierarchical multi-objective optimization balancing fleet minimization with operational cost considerations, and comprehensive energy balance constraints ensuring continuous operation feasibility.

## 3  Problem Description and Formulation for EVRPTW-DWC

The Electric Vehicle Routing Problem with Time Windows and Dynamic Wireless Charging (EVRPTW-DWC) extends the classical EVRP-TW formulation to incorporate both conventional station-based charging and continuous route-based wireless charging capabilities. The problem consists of a set of customers ($I$), charging stations ($F$), and depot locations ($D, E$). The road segments connect pairs of locations; some are equipped with wireless charging infrastructure. Each charging station $f \in F$ can be visited multiple times by vehicles that are used to serve customers. Each electric vehicle has a capacity $C$ and battery limit $Q$. The wireless charging system provides continuous energy transfer with a constant charging rate $w$ along equipped road segments, characterized by coverage fractions $\omega_{ij}$ with $0 \leq \omega_{ij} \leq 1$ that represent partial infrastructure deployment scenarios. If a road segment has no wireless charing capacity, $\omega_{ij}$ is set to 0. We denote $r$ as the energy consumption rate per unit distance, and $g$ as inverse refueling rate at charging station. While we assume only one consumption rate and one refueling rate; multiple refueling rates and consumption rates are considered in the extended version of this study. We assume that vehicles perform full charging at charging stations to restore battery capacity to $Q$, with a fixed charging time of $g \times Q$ per station visit. Each customer $i \in I$ has a demand $q_i$, service time $s_i$, and must be visited exactly once within specified time windows $[a_i, b_i]$.

To obtain a mathematical model, since each charging station is visited multiple times (at most $|I|$ times), we replicate each charging station $|I|$ times, obtain the set $F_{rep}$ of virtual charging stations, and we can assume that each charging station is visited at most once. Let assume $G = (V, A)$ where $V = I \cup F_{rep} \cup \{D, E\}$ denotes all possible locations; and $A \subseteq V \times V$ denote all possible and valid traveling arcs between two locations in $V$. The parameter $s_i$, $[a_i, b_i]$ are set to $0, [0, \infty]$ for $i \notin I$ to simplify the presentation of the formulation.

*Decision Variables*

The formulation employs the following decision variables to model routing decisions, energy management, and temporal constraints:

- $x_{ij} \in \{0,1\}$ ($\forall (i,j) \in A$): indicate whether a vehicle is traveled from location $i$ to location $j$, 0 otherwise.
- $\tau_j \geq 0$ ($\forall j \in V$): arrival time at location $j$ of a vehicle at each location
- $u_j \geq 0$ ($\forall j \in V$): Remaining cargo capacity when leaving location $j$ (units of demand)
- $y_j \geq 0$ ($\forall j \in V$): Remaining battery charge when leaving location $j$ (energy units)

*Objective Function*

The objective function implements a hierarchical optimization approach that prioritizes fleet minimization while secondarily optimizing operational costs, and total travel time and service time:

$$\min \quad M_1 \sum_{i \in I} x_{Di} + M_2 \sum_{(i,j) \in A} d_{ij} x_{ij}$$

$$+ M_3 \left( \sum_{(i,j) \in A} t_{ij} x_{ij} + \sum_{i \in I} s_i \sum_{(j,i) \in A} x_{ji} + gQ \sum_{i \in F_{rep}} \sum_{(j,i) \in A} x_{ji} \right)$$

*Constraint Sets:* The constraints are organized into several logical categories to ensure model comprehensiveness and clarity.

*Customer Service and Flow Conservation Constraints:*

$$\sum_{(i,j) \in A} x_{ij} = 1 \quad \forall j \in I \quad (1)$$

$$\sum_{(i,j) \in A} x_{ij} \leq 1 \quad \forall j \in F_{rep} \qquad (2)$$

$$\sum_{(i,j) \in A} x_{ij} = \sum_{(j,i) \in A} x_{ji} \quad \forall j \in V \setminus \{D, E\} \qquad (3)$$

$$\sum_{(D,j) \in A} x_{Dj} = \sum_{(i,E) \in A} x_{iE} \qquad (4)$$

Constraint [1] ensures each customer receives exactly one visit. Constraint [2] allows each virtual charging station to be visited at most once. Constraint [3] maintains flow conservation. Constraint [4] balances fleet utilization.

*Temporal Feasibility Constraints:*

$$\tau_D = 0 \qquad (5)$$

$$\tau_i + (t_{ij} + s_i)x_{ij} \leq \tau_j + M(1 - x_{ij}) \quad \forall i \in I \cup \{D\}, (i,j) \in A \qquad (6)$$

$$\tau_i + (t_{ij} + gQ)x_{ij} \leq \tau_j + M(1 - x_{ij}) \quad \forall i \in F_{rep}, (i,j) \in A \qquad (7)$$

Constraint [5] establishes the temporal reference point by setting the departure time from the depot to zero for all vehicle routes. Constraint [6] enforces temporal consistency by ensuring that arrival times at destination locations account for both travel time and service time requirements at origin locations, with the big-M formulation activating only when the corresponding arc is traversed. Constraint [7] extends temporal consistency to charging stations, ensuring that departure scheduling reflects the full recharging process with fixed charging time $g \times Q$ when the station is visited.

*Capacity Constraints:*

$$u_j \leq u_i - q_j x_{ij} + C(1 - x_{ij}) \quad \forall i \in V, \forall j \in I \qquad (8)$$

$$u_D = C \qquad (9)$$

Constraint [8] tracks vehicle capacity utilization by reducing the remaining cargo capacity when serving customers with demand $q_j$, while the big-M formulation ensures the constraint remains inactive for non-traversed arcs. Constraint [9] initializes all vehicles with full cargo capacity $C$ at the depot.

*Energy Management Constraints:*

$$y_j \leq y_i + (wd_{ij}\omega_{ij} - rd_{ij})x_{ij} + Q(1 - x_{ij}) \quad \forall i \in I, (i,j) \in A \qquad (10)$$

$$y_j \leq Q + (wd_{ij}\omega_{ij} - rd_{ij})x_{ij} \quad \forall i \in F_{rep} \cup \{D\}, (i,j) \in A \qquad (11)$$

$$y_D = Q \qquad (12)$$

$$y_i = Q \sum_{(j,i) \in A} x_{ji} \quad \forall i \in F_{rep} \quad (13)$$

Constraint [10] manages battery levels during travel between customer locations by accounting for energy consumption at rate $r$ per unit distance while simultaneously capturing wireless charging benefits at rate $w$ along segments with coverage fraction $\omega_{ij}$. The big-M formulation ensures the constraint is active only for traversed arcs. Constraint [11] handles energy dynamics when departing from charging stations or depot locations, where vehicles begin with full battery capacity $Q$ and experience energy changes based on travel requirements and wireless charging opportunities. Constraint [12] ensures that all vehicles commence their operational routes with fully charged batteries at the depot facility. Constraint [13] implements the charging station functionality by setting battery levels to full capacity $Q$ whenever vehicles visit charging facilities.

*Variable Domain Constraints*

$$x_{ij} \in \{0,1\} \quad \forall (i,j) \in A \quad (14)$$

$$\tau_j \in [a_j, b_j] \quad \forall j \in V \quad (15)$$

$$u_j \in [0, C] \quad \forall j \in V \quad (16)$$

$$y_j \in [0, Q] \quad \forall j \in V \quad (17)$$

Constraint [14] defines routing variables $x_{ij}$ as binary, ensuring that arcs between locations are either traversed or not traversed. Constraint [15] restricts arrival times $\tau_j$ to specified time windows $[a_j, b_j]$, enforcing customer service requirements and maintaining temporal feasibility. Constraint [16] bounds remaining cargo capacity $u_j$ within $[0, C]$ to prevent negative capacity and capacity violations. Constraint [17] limits battery charge levels $y_j$ to $[0, Q]$, ensuring vehicles operate within energy constraints and do not exceed maximum battery capacity. These domain restrictions collectively maintain solution feasibility and prevent physically impossible operational states.

## 4 Computational Study

The computational study evaluates the proposed EVRPTW-DWC formulation across two primary objectives:

- Assessing wireless charging infrastructure impact on operational performance.
- Examining the hierarchical optimization approach effectiveness in balancing fleet minimization with operational cost objectives.

Experiments utilize benchmark instances from Schneider et al. [6] with eight instances ranging from 5 to 10 customers, representing small to medium-scale urban delivery scenarios. The optimization model is implemented in Python 3.9 using PuLP with CPLEX 22.1.1, executed on Intel Core i5-13420H with 16 GB RAM. Each instance permits up to two charging station visits with a solver time limit of 3,000 seconds and 0.1% optimality gap tolerance.

The wireless charging system configuration establishes a charging rate of $w = 0.9$ energy units per distance unit, representing conservative estimates based on current dynamic wireless charging technology capabilities. This parameter reflects approximately 90% charging efficiency relative to energy consumption rates, aligning with empirical studies demonstrating that DWC systems achieve 85–95% power transfer efficiency in operational deployments.

The hierarchical objective function employs strategic weighting with vehicle count prioritized at 10,000 units compared to 1.0 unit weighting for both time and distance components. This substantial weight differential ensures that fleet minimization takes absolute precedence in optimization decisions, while secondary objectives receive balanced consideration for operational efficiency improvements.

Wireless charging deployment evaluation encompasses four distinct coverage scenarios designed to reflect realistic infrastructure rollout phases. The baseline scenario with no coverage ($\omega_{ij} = 0\%$) establishes performance benchmarks under conventional charging station dependence. Light coverage ($\omega_{ij} = 20\%$) represents initial pilot deployments targeting high-traffic corridors. Moderate coverage ($\omega_{ij} = 40\%$) corresponds to expanded municipal investments in wireless infrastructure development. High coverage ($\omega_{ij} = 60\%$) reflects mature deployment levels observed in advanced smart city implementations, representing optimal infrastructure investment without excessive capital requirements.

**Results and Analysis**

| Instance | Coverage | Vehicles | Distance | Time | Total Cost | Total % | Solve Time |
|---|---|---|---|---|---|---|---|
| C101C5 | 0% | 2 | 268.1 | 1121.7 | 1389.8 | 0.00% | 0.6s |
|  | 20% | 2 | 252.6 | 956.1 | 1208.7 | -13.03% | 1.1s |
|  | 40% | 2 | 243.2 | 771.3 | 1014.5 | -27.01% | 0.9s |
|  | 60% | 2 | 240.0 | 690.0 | 930.0 | -33.10% | 1.0s |
| C103C5 | 0% | 1 | 184.5 | 1031.8 | 1216.3 | 0.00% | 0.2s |
|  | 20% | 1 | 168.1 | 845.8 | 1013.9 | -16.64% | 0.3s |
|  | 40% | 1 | 168.1 | 795.8 | 963.9 | -20.75% | 0.4s |
|  | 60% | 1 | 164.8 | 614.8 | 779.6 | -35.91% | 0.4s |
| C208C5 | 0% | 1 | 158.5 | 950.1 | 1108.6 | 0.00% | 0.2s |

| Instance | Coverage | Vehicles | Distance | Time | Total Cost | Total % | Solve Time |
|---|---|---|---|---|---|---|---|
| | 20% | 1 | 158.2 | 817.9 | 976.1 | -11.95% | 1.9s |
| | 40% | 1 | 163.0 | 712.6 | 875.6 | -21.02% | 2.0s |
| | 60% | 1 | 157.7 | 607.7 | 765.4 | -30.97% | 1.2s |
| R104C5 | 0% | 2 | 136.7 | 200.5 | 337.2 | 0.00% | 0.8s |
| | 20% | 2 | 136.7 | 198.0 | 334.7 | -0.74% | 5.4s |
| | 40% | 2 | 136.4 | 186.4 | 322.9 | -4.25% | 5.9s |
| | 60% | 1 | 143.4 | 199.8 | 343.3 | +1.81% | 0.5s |
| R105C5 | 0% | 2 | 156.1 | 234.4 | 390.5 | 0.00% | 0.5s |
| | 20% | 2 | 151.1 | 229.0 | 380.2 | -2.64% | 0.6s |
| | 40% | 2 | 151.1 | 201.1 | 352.3 | -9.78% | 0.6s |
| | 60% | 2 | 151.1 | 201.1 | 352.3 | -9.78% | 0.9s |
| C205C10 | 0% | 2 | 229.7 | 1490.4 | 1720.1 | 0.00% | 34.2s |
| | 20% | 2 | 229.7 | 1425.5 | 1655.2 | -3.77% | 51.1s |
| | 40% | 1 | 276.6 | 1573.9 | 1850.5 | +7.58% | 22.4s |
| | 60% | 1 | 271.9 | 1392.0 | 1664.0 | -3.26% | 17.7s |
| C101C10 | 0% | 3 | 397.2 | 2052.2 | 2449.3 | 0.00% | 1473.4s |
| | 20% | 3 | 379.2 | 1702.5 | 2081.7 | -15.00% | 2478.7s |
| | 40% | 3 | 366.4 | 1401.1 | 1767.5 | -27.80% | 3000.0s |
| | 60% | 3 | 378.4 | 1278.4 | 1656.8 | -32.40% | 3000.0s |
| R103C10 | 0% | 2 | 207.1 | 356.3 | 563.4 | 0.00% | 3000.0s |
| | 20% | 2 | 197.2 | 323.2 | 520.4 | -7.60% | 3000.0s |
| | 40% | 2 | 188.7 | 301.5 | 490.2 | -13.00% | 3000.0s |
| | 60% | 2 | 188.7 | 288.7 | 477.3 | -15.30% | 3000.0s |

**Table 1.** Computational Results Summary: Impact of Wireless Coverage on Operational Performance

*Hierarchical Optimization Effectiveness:* The experimental results demonstrate that the hierarchical optimization approach successfully balances fleet minimization with operational cost optimization. The 10,000:1:1 weighting scheme effectively prioritizes fleet reduction while maximizing secondary objective improvements within feasible configurations. Instance R104C5 provides clear validation of this mechanism: at 60% coverage, the optimizer correctly prioritizes fleet reduction from 2 to 1 vehicle despite accepting a 1.81% increase in operational costs, demonstrating that the massive primary objective improvement significantly outweighs modest secondary objective deterioration. Instance C205C10 similarly achieves fleet reduction at 40% coverage while optimizing costs at 60% coverage, confirming the approach's effectiveness in managing competing objectives.

*Wireless Charging Infrastructure Impact:* Statistical analysis reveals progressive operational enhancements across coverage levels, with 60% coverage achieving optimal performance for all instances. Average improvements reach 8.84% ± 5.26% at 20% coverage, 15.35% ± 10.87% at 40% coverage, and 20.36% ± 12.54% at 60% coverage. The fleet reduction analysis demonstrates wireless charging's transformative impact: R104C5 achieves fleet reduction exclusively at 60% coverage, while C205C10 enables reduction at both 40% and 60% levels. Marginal improvements from 20% to 40% coverage average 5.58%, while 40% to 60% improvements average 5.36%, indicating consistent benefit scaling without diminishing returns.

*Route Analysis:* Figure 1 illustrates the practical impact for instance C205C10, where 60% wireless coverage enables fleet reduction from 2 to 1 vehicle while achieving 3.26% cost improvement. Wireless charging provides continuous energy replenishment (gains of +24.3, +13.7, +16.6, and +9.7 energy units along route segments), eliminating charging station dependencies and enabling efficient single-vehicle operation.

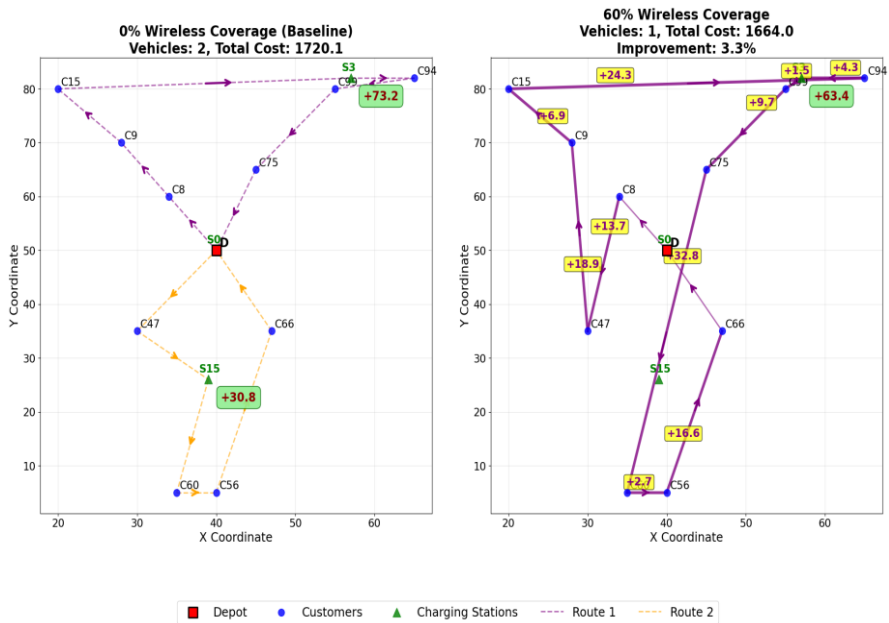

*Figure 1: Route optimization comparison: 0% vs 60% wireless coverage for instance C205C10.*

*Implementation Recommendations:* The evidence supports implementing 60% wireless coverage as the optimal deployment strategy. The hierarchical optimization approach demonstrates consistent effectiveness across all test scenarios, with 60% coverage enabling optimal performance across both fleet efficiency and operational cost objectives. A minimum threshold of 20% coverage provides measurable benefits, while the progression to 60% coverage shows consistent improvement scaling, making it the definitive choice for maximizing operational performance through strategic wireless charging deployment.

## 5   Conclusion and Future Work

This paper presents a comprehensive mathematical framework for the Electric Vehicle Routing Problem with Time Windows and Dynamic Wireless Charging (EVRPTW-DWC), extending Schneider et al.'s foundational EVRP-TW model to incorporate wireless charging capabilities. The key contributions include arc-based dynamic wireless charging representation, partial wireless coverage modeling through parameter $\omega_{ij} \in [0,1]$, and hierarchical multi-objective optimization.

The computational study demonstrates significant operational improvements, with distance and time reductions ranging from 0.7% to 35.9% compared to conventional charging approaches. Critical coverage thresholds emerge supporting infrastructure investment decisions: 20% wireless coverage provides immediate benefits, while 60% coverage offers optimal cost-benefit ratios for infrastructure investment.

The current exact optimization approach effectively handles small to medium-scale instances as demonstrated through computational experiments. However, the MILP formulation's computational complexity limits applicability to larger real-world transportation networks. Future research aims to address this scalability challenge through advanced solution methodologies by using meta-heuristic frameworks.

An important research direction involves developing integrated models for strategic wireless charging lane placement optimization. The current framework assumes predetermined coverage parameters $\omega_{ij}$, but determining optimal infrastructure deployment locations represents a complex bi-level optimization problem. Future research should develop models that simultaneously optimize wireless charging lane placement and vehicle routing decisions, providing transportation planners with systematic decision support tools for maximizing system effectiveness while minimizing capital investment requirements.

## Acknowledgement

The authors (Duc Minh Vu and Ha Minh Hoang) would like to thank the Vietnam Institute for Advanced Study in Mathematics (VIASM) for its kind hospitality and support.